# Counting discrete emission steps from intrinsic localized modes in a quasi-1D antiferromagnetic lattice

M. Sato and A. J. Sievers

Laboratory of Atomic and Solid State Physics and the Cornell Center for Materials

Research

Cornell University, Ithaca, New York 14850-2501

Abstract

**Intrinsic localized modes (ILMs) in a quasi-1D antiferromagnetic material $(C_2H_5NH_3)_2CuCl_4$ are counted by using a novel nonlinear energy magnetometer. The ILMs are produced by driving the uniform spin wave mode unstable with an intense microwave pulse. Subsequently a subset of these ILMs become captured by and locked to a cw driver so that their properties can be examined at a later time with a tunable cw low power probe source. Four wave mixing is used to enhance the emission signal from the few large amplitude ILMs over that associated with the many small amplitude plane wave modes. A discrete step structure observed in the emission signal is identified with individual ILMs becoming unlocked from the driver. At most driver power and frequency settings the resulting emission step structure appears uniformly distributed; however, sometimes, nearby in parameter space, families of emission steps are evident as the driver frequency or power is varied. Two different experimental methods give consistent results for counting individual ILMs. Because of the discreteness in the emission both the size of an ILM and its energy can be estimated from these experiments. For the uniformly** distributed case each ILM extends over ~42 antiferromagnetic unit cells and has an energy value of $1.3 \times 10^{-12}$ [J] while for the case with families the ILM length becomes ~54 antiferromagnetic unit cells with an energy of $1.5 \times 10^{-12}$ [J]. An unresolved puzzle is that the emission step height does not depend on experimental parameters the way classical numerical simulations suggest.

## I. Introduction

Although nonlinear nanoscale localization of energy in atomic lattices was proposed over a decade ago[1,2] and ideas have been put forward for the realization of quantum electromechanical nano-systems[3,4] most work in this area of nonlinear dynamics remains either numerical or theoretical.[5-9] The few experimental demonstrations of this intrinsic nonlinear localization effect have relied on macroscopic lattices to make visual inspection of Intrinsic Localized Modes (ILMs) possible.[10-13] The four exceptions are for: (1) a charge transfer solid PtCl,[14,15] where analysis of resonance Raman spectra was interpreted in terms of ILMs, (2) a quasi-1D antiferromagnetic chain,[16,17] where a high power microwave source was used to produce spin ILMs out of equilibrium, (3) a long lived Amide I band in Myoglobin, which was studied by infrared pump-probe measurements[18] and (4) bcc $^4$He,[19] where inelastic neutron scattering was found to show an anomalous optic-like mode for this monatomic crystal. In all four kinds of experiments direct observation of energy localization has not been possible and observed frequency shifts or time dependences of spectral elements are connected with energy localization by theoretical and/or numerical analysis. Direct measurement of spatial energy localization in such atomic systems is still beyond the frontier, yet these indirect experiments are important for the exploration of non-classical energy localization behavior. Recently we have identified a new ILM signature, namely the discreteness associated with a small number of these excitations in an atomic lattice.[20] As individual ILMs become unlocked from a cw driver they are counted by means of a novel experimental approach. The discreteness appears in the time dependent emission spectra of a quasi-1D antiferromagnet when the system first is driven far from equilibrium and then is subjected to a four wave mixing experiment at an ILM frequency, locked to a cw oscillator set below the lowest Antiferromagnetic Resonance (AFMR).

In the experimental study reported here we explore these localized excitation signatures in more detail. With the ILMs synchronized to a cw driver and for times greater than the spin lattice relaxation time, $T_1$ of the AFMR, the stepwise decrease of the emission strength is measured as a function of different experimental parameters. It is important to have the sample shape such that the AFMR is at the bottom of the spin wave manifold so that when the uniform mode is driven unstable ILMs can appear in the frequency gap. For a fixed cw locking driver input (power and frequency) the integral relation between these emission steps is counted. Interrupting the cw driver with specific time delays is observed to produce multiple step cascades. We confirm the findings in Ref. [20] that the measured emission step height is only weakly dependent on the cw power or on the driver frequency used in the four wave mixing experiment. The experiments reported here suggest that the observed uniform emission steps are outside the expected behavior of a simple locked classical anharmonic oscillator model.

In the next section the properties of the quasi-1-D antiferromagnet are reviewed in order to characterize the spin wave dispersion curves. Next the measurement technique is described in some detail. Section III presents (1) the experimental results for the absorption and nonlinear emission spectra, (2) the time dependent emission from ILMs, and (3) the dependence of the emission strength on the interruption time for the locking driver. The four wave emission results are discussed in Section IV where possible causes of the observed steps in the emission signal are presented. Next the four wave mixing power expected from the uniform AFMR mode is related to the experimental quantities.

A detailed description of the nonlinear mixing magnetization for the AFMR is presented in Appendix (A). Then the same type of derivation is carried out for ILMs to show that the number of ILMs in the sample varies as the square root of the emission signal. After an expression for the stored energy in a driven mode is obtained in Appendix (B) the energy of an ILM is estimated from the data. Finally the step height dependence on driver frequency is calculated for this classical biaxial spin model and it is shown to vary more rapidly than observed for the measured step data. The summary and conclusions follow.

## II. Experimental Details

### A. Antiferromagnetic resonance sample geometry

It has been shown that the localization strength of an ILM in an antiferromagnet is determined by the ratio of the anisotropy field to the exchange field with strong localization occurring for a ratio of order ~1.[7] Below a Néel temperature $T_N = 10.2\ K$ for $(C_2H_5NH_3)_2CuCl_4$ the spin 1/2 $Cu^{2+}$ ions are oriented along the a-crystal axis, in alternating sheets of strong ferromagnetically coupled spins with a weak antiferromagnetic coupling between adjacent sheets[21,22] as illustrated in Fig. 1. At 1.4 K the interlayer antiferromagnetic exchange field is $H'_E = 829\,\text{Oe}$ and the intralayer ferromagnetic exchange field exchange field $H_E = 5.5 \times 10^5\,\text{Oe}$ so that $H'_E/H_E = 1.51 \times 10^{-3}$. In this low temperature region the spins in a given layer are strongly aligned in the same direction so to a good approximation the low frequency spin dynamics can be modeled by a 1-D two sublattice antiferromagnet with each layer represented by a single classical spin. Due to the resulting biaxial anisotropy and the weak antiferromagnetic interaction between these spins the upper and lower frequency uniform modes are polarized along b- and c-crystal axes, respectively. The resulting small amplitude spin-wave dispersion curves are presented in Fig. 2. Figure 2 (a) shows that both the uniform mode frequencies and the bandwidths of the excitation branches along c-crystal axis spin wave directions are in the GHz range. The figure insert identifies the two sublattice lowest frequency AFMR mode with a linearly polarized transverse ac moment generated in the 1-3 direction but not in the 2-4 direction.[23]

There are two important features that need to be added to the picture to characterize the actual spin dynamics. Like all 3-D systems near the zone center the spin wave frequency depends on the angle between the ac polarization direction and the spin wave propagation direction. Spin waves that propagate perpendicular to the polarization direction have the lowest frequency. This angular dependence can be seen in Fig. 2(b). In addition, the 3-D long-range magnetic dipole-dipole interaction creates magnetostatic modes near the zone center so that, unlike the finite wavevector spin waves, the resulting uniform mode and nearby magnetostatic mode frequencies depends on the actual sample shape. The range of possibilities is shown in Fig. 2(b). The frequency can be varied from the bottom of the k//a or k//b spin wave band, up to the bottom of the k//c spin wave band. Since the lower branch has the ac polarization along the c-axis, a plate oriented perpendicular to the c-axis has the highest AFMR frequency while a rod or plate directed along the c-axis has the lowest resonant frequency. Spin-spin relaxation of the uniform mode occurs via the spin states that are degenerate in frequency so only the rod-like sample has a true energy gap below the uniform mode frequency. In order to start with a uniform mode at the bottom of the spin wave manifold mainly samples with rod-like demagnetizing factors are studied here. The linear excitation properties of this system have been investigated earlier by others[24].

The samples, for our nonlinear experiments, are grown from aqueous solution and then cut into a rod-like shapes directed along the c axis, parallel to the polarization of the low frequency uniform mode. For microwave non-resonant ac coupling to the uniform mode two single-turn coils surrounded the sample, which is immersed in superfluid helium and maintained at 1.2 K.

### B. Nonlinear measurement techniques

Our earlier method to generate ILMs in antiferromagnets employed a short, high intensity, microwave pulse to produce a large amplitude instability in the uniform mode (AFMR). An absorption method was then used to examine the time dependent products generated and the eventual recovery of the uniform mode.[16,17,25,26] Our more recent experimental studies of ILMs have explored the uniform mode instability of classical nonlinear micromechanical arrays and demonstrated that, after ILMs are produced, a modest amplitude cw driver can be locked to some modes, resulting in coherent ILMs with fixed amplitudes as long as the driver remains on.[12,27] In Ref. [20] and the experiments described here we apply the same locking idea to an antiferromagnet. First the AFMR mode is driven into the unstable amplitude range with a high power pump, next a lower power, lower frequency, cw driver is applied to lock a subset of the ILMs and finally the product spectrum is then probed with a third oscillator of variable frequency.

Since the third order nonlinearity $\chi^{(3)}$ of the antiferromagnet makes possible a four wave mixing experiment[28-30] this method has been used to observe in emission the small number of ILMs that remain locked to the driver. The transverse magnetization components oscillating at the different frequencies in the four wave mixing process are illustrated in Fig. 3 where both the input and output are identified by the arrow directions.

Figure 4 shows the experimental setup for generation, locking, absorption and four wave mixing experiments in the antiferromagnet. Inside the dotted box is approximately the same setup used in previous absorption measurements.[17] The high power pulse source $f_1$ is fed into a two-turn coil surrounding the sample via a hybrid coupler to generate the ILMs. The absorption spectrum is measured with the low power $f_3$ signal, which is reflected from the sample and sent to the spectrum analyzer and digital recorder. To measure the four-wave mixing emission signal, again both $f_3$ and the spectrum analyzer filter frequency $f_{sp}$ are scanned in but now the condition for the spectrum analyzer is $2f_2 - f_3 = f_{sp}$.

The three sources, $f_1$, $f_2$, and $f_3$ are combined by -10 dB and -20 dB couplers, then fed into the coil via a hybrid coupler. The middle power source $f_2$, a switch labeled SW in Fig. 4, and an amplifier is used to lock the ILMs. Since the reflection of the $f_2$ signal from the sample may exceed the maximum linear input level of the spectrum analyzer and produce a 2nd order nonlinear mixing signal inside of the spectrum analyzer which in turn generates a very weak spurious signal at the 3rd order mixing frequency by another 2nd-order mixing process, part of the $f_2$ driver is summed before the signal goes into the spectrum analyzer and used to cancel the $f_2$ component. This canceling loop works only for the signal starting from the $f_2$ driver and does not affect the emission signal from the sample, owing to the large loss induced by the hybrid isolation and variable attenuator. A diode switch (SW) at the front of the spectrum analyzer is used to block the high-power $f_1$ signal. Since no sharply peaked emission signal is expected if there are no locked ILMs, the entire detectivity range of the spectrum analyzer (-90 dBm to 0 dBm) is available.

## III. Experimental Results

### A. Absorption and emission spectra

Figure 5 shows the dynamic time dependent absorption spectrum for the uniform mode driven to large amplitude with the important microwave frequencies for the experiment identified. The strong microwave pulse $f_1$ drives the uniform mode (AFMR) into an unstable amplitude region where it breaks up into ILMs that initially extend over a large, low frequency interval. The cw locking oscillator $f_2$ picks out a subset of ILMs from this spectrum since its frequency is set at the lower edge of the broad absorption band initially produced by the $f_1$ pulse. The time dependent absorption spectrum shown here is measured by using a weak probe signal $f_3$, which can be tuned over the entire frequency region of interest. To measure an absorption spectrum, both $f_3$ and $f_{sp}$ are scanned in tandem keeping the condition $f_3 = f_{sp}$. In Figure 5 the uniform mode reforms at times somewhat shorter than $T_1$, the spin lattice relaxation time, and approaches its equilibrium configuration at longer times. Note that the number of ILMs that remain locked to the frequency $f_2$ at these long times is, in general, too small to be seen with this absorption technique.

The time dependent emission signal, over the same Fig. 5 time scale, is presented in Fig. 6. The time sequence for the $f_2$ driver is represented by the solid line at the bottom of Fig. 6(a). To reduce the large $f_2$ signal in this low-resolution study two spectra are obtained, one with $f_1$ applied and another without. These are then subtracted to clearly see the emission. Initially, there is no emission signal until the $f_1$ pump pulse is initiated at 20 $\mu s$. This demonstrates that the $f_2$ driver, by itself, cannot strongly excite the AFMR mode. The emission signal only occurs after the $f_1$ pulse creates ILMs. When the $f_2$ driver is turned off, the emission signal decreases immediately and does not completely reappear when the driver is turned on again. Figure 6(b) shows similar time dependent emission results when the $f_2$ driver is interrupted sequentially, demonstrating that unlocking and re-locking of ILMs are well controlled by the switching of the driver.

All of these emission results can be understood in terms of the locking effect. As time progresses the AFMR approaches the equilibrium frequency as illustrated in Fig. 5. At the same time the $f_2$ driver frequency is kept fixed. Turning off the $f_2$ driver releases locked ILMs and allows their phase and frequency to vary. Restarting the $f_2$ driver captures only some of the ILMs, those nearby in phase and frequency. It is important to remember that there are no ILMs generated by the $f_1$ pump excitation at this late stage so turning off the $f_2$ driver significantly disturbs ILM locking. These experiments show that the locked ILM state can be maintained as long as the driver is on so that experimental measurements can now be carried out over a greatly expanded time scale. With the result that the locked ILM state permits a measurement of the emission spectra at high frequency resolution.

Figure 7 shows both absorption and emission spectra obtained at 2 ms after the $f_1$ pulse. (This time interval corresponds to about 3 million periods.) Figure 7(a) has the $f_2$ driver closer to the AFMR at lower power while Fig. 7(b) has the $f_2$ driver farther from the AFMR at a higher power to maintain the locked ILMs. The absorption spectra (dashed) are relatively simple while the more sensitive emission spectra are complex. The strong peak in the absorption spectra (linear scale) is associated with the AFMR mode. Its frequency is pulled down slightly by the presence of the $f_2$ driver. In Fig. 7(a), a weak peak can be seen below the $f_2$ frequency, probably associated with a large number of locked ILMs.

The emission spectra in Fig. 7(a) and (b) (solid line) are displayed using a log scale. The strongest emission peak observed slightly lower than the $f_2$ driver is from the ILMs as previously identified in Fig. 6. The richness of these emission spectra, which will be discussed in Section IV.A, demonstrates that this nonlinear emission method is much more suitable for the detection of a nonlinear species, such as ILMs, than is the linear absorption technique.

**B. Time dependent emission from locked ILMs**

For time dependent emission experiments the detector frequency is now fixed and set at the frequency of the ILM emission maximum. Figure 8 shows the long time decay of the emission signal for a cube-shaped sample where the AFMR is within the spin wave band. There are two contributions to the observed time dependent emission signal as the power is varied in increments. (1) As expected the signal from the ILMs decreases with time because of the increasing frequency difference between the $f_2$ driver and the AFMR, which tends to unlock ILMs for each trace with constant $f_2$ power. But unexpected is the incipient step like structure seen in some traces. (2) The background signal from the AFMR also decreases since the frequency gap to the $f_2$ driver is increasing as illustrated in Fig. 5. The resulting time dependent emission spectra are quite complex although they are repeatable from shot to shot. Only averaged traces are shown here. The use of the square root of the emission on the ordinate is justified in Section IV. C.

For the next set of experiments the sample is a c-axis directed thin plate, which has the same demagnetizing factor as a rod so that the linearly polarized uniform mode is now at the bottom of the spin wave spectrum. The traces in Fig. 9(a) correspond to a fixed driver frequency with its power settings varied in increments. In Fig. 9(b) the power is kept fixed and the frequency of the driver is varied in increments. Surprisingly, a distinctive and reproducible step structure is observed, with steps of similar height, when the square root of the emission power is plotted. Apparently individual ILMs are being counted. At most other power and frequency settings the data traces are similar to those displayed in Fig. 9(a, b); however, sometimes, nearby in parameter space, a clearly different structure appears in the data as the driver frequency or power is varied. Figures 9(c, d) show the emission results for such cases. The actual traces are similar to but not identical with those shown in Figs. 9(a, b) but gaps now appear between some traces breaking them up into families while that clear distinction is not evident in Figs 9(a, b). Similar results are obtained for the c-axis directed rod sample.

Compared to the observed complex behavior of time dependent signal for the cube shape the c-axis directed thin plate and rod samples both show smooth, well-determined emission step decays as a function of $f_2$ frequency and power. This clear difference between the results in Figs. 8 and 9 convince us that it is necessary to have the AFMR near the bottom of the spin wave manifold to observe locked ILMs. Apparently the interaction between the ILMs and the manifold states in a cube or c-axis directed perpendicular to a flat plate shaped sample like those used in Ref. [16] does not permit stable locked ILMs.

Figure 10 demonstrates how individual emission decay curves can be decomposed into a sum of steps of equal height plus an exponential decay when the square root of the emission power is plotted on the ordinate. The dotted curves in Figs. 10(a) and 10(b) are the same dotted curves in Fig. 9(a) and 9(b), respectively. The solid smooth curves in each frame represent an exponential curve displaced by multiples of a specific offset. Each of the decay curves matches one of the solid curves between increments. The emission step height, $\Delta\sqrt{P^{(3)}}$, in Fig. 10 is evaluated from the constant offset. The measured values given in Table I for uniformly distributed steps in Fig. 9(b) and for steps in Fig. 9(d) where steps are divided into groups by a gap

## C. Dependence of the emission on the $f_2$ locking time delay

The time delay locking results shown in Fig. 6 suggest another kind of unlocking experiment with which to examine the step structure. Due to the eternal behavior of locked ILMs, the emission signal can be recorded at very long times when the uniform mode back ground emission signal has nearly vanished. The dependence of emission decay curve on a locking time delay of the $f_2$ driver is shown in Fig. 11. In these experiments the driver is turned off at 2 ms for a short interval and the emission is recorded as this delay interval is varied. The dotted and dash-dotted curves are for delay times of 10 and 50 $\mu s$. For the dot-dashed curve the off and on time of the driver is shown. The emission at times beyond 2 ms depends on this delay time hence the longer the delay the smaller the emission and hence the smaller the number of locked ILMs remaining. The experiment then is to measure the emission at 4 ms as a function of the locking delay time at 2 ms.

A summary of the experimental results is presented in Fig. 12. The square root of the emission power at t = 4 ms is shown in Figs. 12(a, b) for a vertical plate sample. Panel (a) has the higher frequency. The solid trace is for one driver power and the dot dashed trace for a higher driver power. The horizontal dashed markers are guides to the eye, with the distance between them adjusted to fit the solid experimental data steps. This distance is the same in frames (a) and (b). The results presented in Fig. 12(c) show two power values at one driver frequency for a rod shaped sample. The step scale is now different. The data mostly show vertical steps between horizontal regions associated with different ILM numbers indicating that the ILM emission appears on a constant background. The higher power cases show almost the same step height, although the overall traces are shifted upwards. This shift is in emission is due to an unknown source. An interesting observation is that often simultaneous double steps are observed at lower signal levels. A fractional step is seen in panel (b), 2nd step from the top, where the height is half of the usual step size. This height is probably also included in the large, 4th step from the top of the higher power case (dotted curve), since the 2nd level from the top and the lowest level are almost the same.

## IV. Discussion

### A. Four wave emission spectrum

The emission signal is observed only when the probe frequency and the spectrum analyzer frequency are chosen to satisfy the relation $f_{sp} = 2f_2 - f_3$. The resulting emission spectrum is as sharp as the probe oscillator line width if the probe frequency is fixed and only the spectrum analyzer frequency is scanned. Thus, the emission signal is due to the 3rd order mixing of signals, which are time-coherently induced in the sample by the monochromatic driver. No emission signals are observed from incoherent spin fluctuation, which may exist around the locking driver frequency $f_2$. Such signals would be below the noise level. Since the spectra shown in Figs. 7(a) and (b) are obtained by scanning the probe frequency $f_3$, the structured spectra are caused by the response to $f_3$ stimulation, and not induced by $f_2$. In other words, the emission structure is the result of resonances of $f_3$ or the $2f_2 - f_3$ mixing signal with some resonator. Simply stated, the sample contains one 3rd order mixer and two kinds of resonators, namely, the ILM and the uniform mode (AFMR).

The broad emission peak at the AFMR frequency in Figs. 7(a) and (b) is the four-wave mixing emission from the uniform mode. The $f_3$ probe resonates with the AFMR and the resulting magnetization oscillates at the frequency $f_{sp} = 2f_2 - f_{AFMR}$. As described in Appendix A, another peak is expected at $f_3 = 2f_2 - f_{AFMR}$ and such structure is observed. In this case, the $f_3$ probe is converted to a magnetization at $f_{sp} = f_{AFMR}$, which then resonates with the AFMR. The shoulder observed at 1.32 GHz in Fig. 7(a), the weak broad peak at 1.28 GHz in Fig. 7(b) and a shoulder in Fig. 7(a) are all associated with this process. In general, for each resonator, two sideband peaks are observed on both sides of the locking frequency. The two peaks are due to the probe resonance or to the converted frequency resonance. If the nonlinear response function with fixed $f_2$ is proportional to the product of two linear response functions as given by Eq. (A13) in Appendix A then the two sidebands will have equal height[28]. However, in our experiments the process is more complex since the probe resonating case has the stronger response.

The strongest emission peak in the spectrum shown in both Figs. 7(a) and 7(b) is just to the lower sides of $f_2$ frequency. It is emission from the locked ILMs. A somewhat weaker emission response appears on the other side of the driver so these two peaks are the side band pair for the ILMs. The existence of such side band structure in the response spectrum has been identified in simulations with ILMs in the antiferromagnet[26] and also for a micromechanical oscillator systems.[31] In our experimental case, the ILM resonant frequency should be lower than the driver frequency; hence the emission frequency is pulled down. The strong low frequency peak is the probe resonating signal and it has larger amplitude than the converted signal resonance. The much weaker background emission seen over the entire frequency interval in Fig. 7 is not been characterized.

It should also be mentioned that in nonlinear optics, cross phase modulation is larger than self phase modulation, i.e., the frequency shift of the small amplitude mode due to the nonlinear excitation (cross frequency shift) is much larger than the frequency shift of the nonlinear (large amplitude) mode due to its large amplitude (self frequency shift)[32] thus two sidebands can be expected for resonance locking for a variety of nonlinear systems.

**B. Steps in emission**

*1. As a function of the AFMR recovery*

For a qualitative discussion of these experimental emission step results it is helpful to focus on what might be expected to occur for a simple classical ILM oscillator system. The emission peaks in Fig. 7 demonstrate that ILMs are locked to the $f_2$ driver. The frequency difference between the AFMR and the $f_2$ driver, $\Delta f = f_{AFMR} - f_2$, increases with time in the emission step observations. The nonlinear resonance response versus frequency curve for a classical anharmonic oscillator, shown in Fig. 13(a), can be used to illustrate how steps could appear. Experimentally an emission step is observed after the experimental parameters are adjusted so that at a particular time after the $f_1$ pulse the set of locked ILMs are at a $\Delta f$ amplitude value just below the peak represented by the open circle in Fig. 13(a). As time progresses $\Delta f$ increases, the open circle moves to the right, and finally reaches the point where the amplitude switches from the large value to the small one, the ILM excitation becomes unlocked from the driver, and the emission decreases suddenly. This description would account for one step but does not explain a series of emission steps of equal height like those observed experimentally. Below we consider four possible explanations for such an emission ladder.

*i.* Steps produced by sample inhomogeneity: The individual ILMs would now have different amplitudes at $f_2$ so that the unlocking times would occur at different $\Delta f$ values giving multiple steps; however, both the observed step heights and the time intervals between steps are regular, not a distribution as one would expect for this sort of process.

*ii.* Steps produced by antiferromagnetic domains: No difference in the emission step structure was observed in experiments shown in Fig. 9(a, b) between the initial state when the sample was cooled in zero magnetic field and the final zero-magnetic field state obtained by cycling a magnetic field directed along the easy crystal (a) axis through the spin flop field value and back to zero.

*iii.* Steps produced by magnetostatic modes: These modes are not predicted to occur below the AFMR frequency for a rod shaped sample, but even if such resonances could participate in the emission signal formation, the resulting steps would not be of equal height.

*iv.* ILMs coupled by spin waves: Our final classical explanation of the equal step height observation is based on the premise that each ILM is identical (locked) and that these few localized 1-D excitations are coupled to each other via spin waves acting as intermediate states. The form of the resulting ferromagnetic coupling between ILMs is assumed similar to that proposed by Suhl[33] and Nakamura[34] for the coupling between nuclear spins via spin waves in antiferromagnets. The resultant interaction resembles a screened $1/r$ potential. Two differences are expected for this case under discussion: (1) the ILM excitation frequency is very close to the spin wave band frequencies so the resulting screening length is larger than the sample size and (2) since these ILMs are made up of 2-D sheets of aligned $Cu^{2+}$ spins, a divergence theorem argument suggests that the interaction between the 1-D ILMs would be independent of the distance between them. The ILM frequency $f_\ell$ should now depend on the small number $n$ of ILMs in the lattice so that the shifted ILM frequency

$$f_{\ell,n} = f_{\ell}(1 + bn), \tag{1}$$

where $b$ is the dimensionless ILM-ILM coupling constant. Now the experimental parameters are adjusted so that at a particular time after the $f_1$ pulse the set of locked ILMs are at a $\Delta f$ amplitude value (open circle) just below the peak in Fig. 13(b). Again for this experiment since $\Delta f$, the interval between the set of ILMs locked to the driver and the relaxing AFMR frequency increases with time, the open circle moves to the right. When it reaches the point where the amplitude switches from one value to the next smaller one, one ILM excitation becomes unlocked from the driver, the emission decreases suddenly, the remaining ILMs continue to remain locked but now with a smaller total amplitude and, by Eq. (1), with a shifted switching frequency. The resulting stepped emission signature would now appear as individual ILMs become unlocked, one by one as shown in Fig. 13(b). There is one remaining difficulty with this explanation. Classically the ILM amplitude, i.e., emission step height, should depend on $\Delta f$ as we show below in Section (IV. E); however, the observed emission steps display a remarkably constant step height, almost independent of this frequency gap, as determined in Fig. 9(b, d).

## 2. As a function of locking driver time delay

The data for the second kind of step experiment giving the results shown in Fig. 12 permits one to examine the emission step production in a different way. First the experimental parameters are adjusted so that the ILM emission is obtained as represented by the open circle on the amplitude plot in Fig. 13(b) then since $\Delta f$ is now fixed, the earlier time delay simply changes the number of ILMs that are locked so the open circle now moves in the vertical direction. If the emission varied continuously with time delay then the data would appear as a straight line on this semilog plot; however, the observed emission value decreases in units for a time delay that increases continuously. The fact that steps and multiple steps appear in Fig. 12 is not too surprising since the driver time delay for fixed $\Delta f$ does not have the selectivity of the time dependent $\Delta f$ variation experiments shown in Fig. 9.

Another interesting feature is the fractional step height shown in Fig. 12 (b). This signature may come from an ILM captured by a trapping site such as an impurity in the sample. If the captured state were energetically favored for the ILM, the step height would be smaller than for a free ILMs, since a trapped state has a smaller transverse moment.

## C. Form of the four wave mixing equation for an antiferromagnet

The change in four wave mixing magnetization associated with a step and its relation to the change in magnetization of an ILM is now estimated. First we consider the AFMR four-wave emission. The voltage $V$ at the coupling coil is directly related to the changing four wave mixing transverse magnetization by

$$\hat{V} = -\frac{d\hat{\Phi}}{dt} = \omega\mu_0 vk\hat{M}_c^{(3)} \tag{2}$$

where $\Phi$ is the flux through the coil, changing at frequency $\omega$, $\mu_0$, the permeability of free space and $\hat{M}_c^{(3)}$ is the third order, total nonlinear magnetization in the c-direction. In our experiments only a c-axis directed plate sample (rod demagnetization factor for the low frequency mode) is analyzed in detail here. Its volume $v = 2.25 \times 2.8 \times 0.8 mm^3$ and the field-current constant for the two single loop coil, $k = 180 [m^{-1}]$. The resulting power at the detector is

$$P^{(3)} = \frac{1}{2} \frac{|\hat{V}|^2}{R} = \frac{1}{2} \frac{(2\omega_2 - \omega_3)^2}{R} \mu_0^2 v^2 k^2 |M_c^{(3)}|^2, \tag{3}$$

where the detector impedance, $R = 50\Omega$.

Following Refs. [29,30] an approximate scalar expression for the nonlinear transverse third order magnetization for the uniaxial case associated with the four wave mixing process is outlined in Appendix A. In analogy with Eq. (A14), which gives the third order nonlinear transverse magnetization in terms of the linear transverse magnetizations at the driver and probe frequencies for a uniaxial antiferromagnet, the frequency dependent nonlinear transverse magnetization of the lowest uniform mode along the c direction is written as

$$\hat{M}_c^{(3)}(2\omega_2 - \omega_3) = \frac{3}{64} \frac{1}{M_0^2} \frac{\hat{\chi}_c(2\omega_2 - \omega_3)}{\chi_\perp(0)} [\hat{M}_c(\omega_2)]^2 \hat{M}^*_c(\omega_3), \tag{4}$$

where the sublattice magnetization is $M_0$, $\chi_\perp(0)$ is the low temperature dc susceptibility[35] contributed by the c axis-polarized uniform mode. Here the transverse magnetization at frequency $\omega_i$ is

$$\hat{M}_c(\omega_i) = \hat{\chi}_c(\omega_i) \hat{H}_c(\omega_i). \tag{5}$$

The emitted power $P^{(3)}$ is proportional to square of the 3rd order nonlinear ac magnetization $\hat{M}_c^{(3)}(2\omega_2 - \omega_3)$ so

$$\sqrt{P^{(3)}} = \left(\frac{\alpha}{2R}\right)^{1/2} \left|\hat{M}_c^{(3)}\right| \mu_0 vk(2\omega_2 - \omega_3)$$
$$= \frac{3}{64}\left(\frac{\alpha}{2R}\right)^{1/2} \frac{(2\omega_2 - \omega_3)\mu_0 vk}{M_0^2} \frac{\left|\hat{\chi}_c(2\omega_2 - \omega_3)\right|}{\chi_\perp(0)} \left|\hat{M}_c(\omega_2)\right|^2 \left|\hat{M}_c(\omega_3)\right| \quad (6)$$

where a calibration factor $\alpha$ has been introduced to account for differences between the uniaxial and biaxial cases. This factor is then evaluated by measuring the emission signal obtained for the AFMR value. The observed emission feature at 1.28 GHz shown in Fig. 7(b) is associated with the bulk four wave mixing signal when $f_3$ crosses the uniform AFMR value.

To determine the value of $\alpha$ the measured parameters in Eq. (6) have been determined and are given in Table I, column 2. An additional necessary quantity is the sublattice magnetization, $M_0 = 1.78 \times 10^4 \, A/m$. A standard Lorentz oscillator form is assumed for the dynamic susceptibility so

$$\hat{\chi}_c(\omega) = \frac{\omega \omega_0 \chi_\perp(0)}{\omega_0^2 - \omega^2 + i\gamma\omega}, \quad (7)$$

where $\omega_0$ is the AFMR resonance frequency, $\gamma/2\pi$ is the measured linewidth, and $\chi_\perp(0) = 0.14$. Comparing the calculated value for $\left|\hat{M}_c^{(3)}\right|$ with the experimental value gives $\alpha = 31$.

**D. Step emission from four wave mixing**

When the nonlinear excitations are ILMs, $\hat{M}_c^{(3)}(2\omega_2 - \omega_3)$ is calculated by summing the individual ac magnetic moments and dividing by the sample volume so that

$$\hat{M}_c^{(3)}(2\omega_2 - \omega_3) = \frac{nv_l \hat{M}_{lc}^{(3)}(2\omega_2 - \omega_3)}{v}, \tag{8}$$

where $n$ is the number of ILM, $v$ is the volume of the sample, $v_l$ is the volume of one ILM, and $M_{lc}^{(3)}$ is the nonlinear transverse magnetization along c axis for an ILM. Since each locked ILM responds the same way to the $f_2$ driver and the $f_3$ probe it contributes the same amount to the net nonlinear transverse magnetization.

To calculate the step emission power the first assumption is that Eq. (4) remains valid for an individual ILM so that

$$\hat{M}_{lc}^{(3)}(2\omega_2 - \omega_3) = \frac{3}{64} \frac{1}{M_0^2} \frac{\hat{\chi}_{lc}(2\omega_2 - \omega_3)}{\chi_\perp(0)} \left[\hat{M}_{lc}(\omega_2)\right]^2 \hat{M}^*_{lc}(\omega_3). \tag{9}$$

Substituting Eq. (9) into Eq. (8) and taking the modulus gives the desired power expression, namely,

$$\sqrt{P^{(3)}} = \frac{3}{64}\left(\frac{\alpha}{2R}\right)^{1/2} \frac{(2\omega_2 - \omega_3)\mu_0 vk}{M_0^2} \frac{|\hat{\chi}_{lc}(2\omega_2 - \omega_3)|}{\chi_\perp(0)} n\left(\frac{v_\ell}{v}\right) |\hat{M}_{lc}(\omega_2)|^2 |\hat{M}_{lc}(\omega_3)|, \tag{10}$$

where the calibration factor $\alpha$ is assumed to have the same value (= 31) measured for the four wave mixing for the AFMR described above and presented in Table I. Thus an emission step observed in the square root of the emission power is associated with the disappearance of one ILM.

The measured emission values for a single step $\Delta\sqrt{P^{(3)}}$ given in Table I for the uniformily distributed cstep and for the family distributed step can be used in Eq. (10) to estimate the volume fill fraction of an ILM with the added condition that now the ILM transverse magnetization is fully resonant with the driver and nonlinear. Since the sum rule strength of an oscillator is independent of its nonlinearity[36] we approximate the ILM susceptibility peak by the Lorentz oscillator value: $\chi_\perp(0)\omega_\ell/\gamma$. For the linear response function $\hat{\chi}_{\ell c}(2\omega_2 - \omega_3)$ and the linear magnetization $\hat{M}_{\ell c}(\omega_3) = \hat{\chi}_{\ell c}(\omega_3)\hat{H}(\omega_3)$ in Eq. (10), we use Eq. (7) with the resonance frequency $\omega_0$ now set at the peak emission frequency. The other parameters in Table I can then be used to estimate the volume fill fraction per ILM,

$$\frac{v_\ell}{v} = \frac{\Delta L}{L}. \tag{11}$$

For the sheet geometry of aligned spins in the 1-D antiferromagnet where $\Delta L$ is the spatial length of an ILM, and $L$ is the sample length, the estimated ILM lengths for the two different cases are given in Table I in terms of antiferromagnetic unit cells. For comparison classical MD simulations with a model 1-D antiferromagnetic system give a somewhat smaller ILM length of 7 antiferromagnetic unit cells. Finally the number of spins in an ILM $N_{spin}$ is also presented in Table I. (This number is required for calculating the energy per spin in the next section.)

**E. Energy estimate for an ILM**

Since the steps are well defined there is some value in estimating the ILM energy. The height of one step $\Delta(P^{(3)})^{1/2}$ can be directly related to $|\hat{M}_{\ell c}(\omega_2)|^2$ of a single ILM and hence to the ILM energy. The stored energy density expression for a driven uniform mode is developed in Appendix B and given by Eq. (B6). Integrating this expression over the sample volume gives

$$\Delta E(M_c) = \int \frac{\mu_0 |\hat{M}_c|^2}{4 M_0} H_{c-eff} dv, \qquad (12)$$

where $H_{c-eff}$ is the effective internal field for resonance and $|\hat{M}_c|$ is the modulus of the driven magnetization.

Next we assume that Eq. (12) remains valid for driven ILMs. From our experimental results only a few locked localized modes need to be added to the uniform mode. Integrating over the sample volume gives the factor $nv_l$ since only the ILMs are driven so the AFMR contribution on each side of the step cancels. The energy change associated with a single step ($\Delta n = 1$) is

$$\delta(\Delta E_c) = \frac{\mu_0 H_{c-eff} v_l |\hat{M}_{\ell c}(\omega_2)|^2}{4 M_0}. \qquad (13)$$

Substituting the experimental values into Eq. (13) gives the desired energy: $\delta(\Delta E_c)$ or the energy per $Cu^{2+}$ spin: $\delta(\Delta E_c)/N_{spin}$. The values for the two types of data described in Fig. 9(b) and 9(d) are given in Table I.

**F. Calculated step height dependence on frequency and power**

From classical numerical simulations for the biaxial material as $\Delta\omega$ increases the height of an ILM increases and its width decreases. Since each ILM spin contributes to the mixing signal, the step height is proportional to the Sum of the Square of the Transverse Component (SSTC) of each spin

$$\Delta\sqrt{P^{(3)}} \propto \sum_n |\hat{s}_{\ell c,n}|^2 = SSTC, \tag{14}$$

where $\hat{s}_{\ell c,n}$ is the complex amplitude of the n$^{th}$ spin of the ILM. The eigenvector of an ILM is calculated as in Refs. [37,38] and the dependence of the SSTC versus normalized gap frequency is presented in Fig. 14. This calculated dependence should be contrasted with the experimentally measured emission step heights shown in Fig. 9(b). Expanding these data we find the values are nearly the same when $\Delta f$ is scanned at a fixed power level. For the large number of steps in the central region of this figure the observed change is less than 4% for the frequency range 1.325 - 1.333GHz ($\Delta\omega/\omega = 0.0257 \leftrightarrow 0.0199$), while the simulation results (Fig. 14) indicate a 12 % increase with $\Delta\omega/\omega$ covering the same range. For Fig. 9(d) the same kind of analysis gives less than 3% for the frequency range 1.330 - 1.338 GHz ($\Delta\omega/\omega = 0.0221 \leftrightarrow 0.0162$), while now the simulation results indicate an 15 % increase. This behavior cannot be explained with our classical eigenvector calculation.

## V. Summary and Conclusions

Countable ILMs have been observed in an atomic lattice with a nonlinear energy magnetometer. The instrument first produces frequency locked ILMs in a quasi-1D antiferromagnet and then measures the four wave mixing signal emitted by the sample versus time or versus locking driver delay time. This technique makes observable in nonlinear emission the small number of ILMs that remain in steady state. The stabilization of these locked ILMs makes possible their spectroscopic study at high resolution. Because these excitations are strongly nonlinear, four-wave mixing emission spectroscopy is an ideal way to enhance the ILM signal over that obtained from the more numerous plane wave spin excitations. This magnetometer technique is much more sensitive than the absorption technique previously used.

Emission step structure has been found as a function of the AFMR recovery time and as a function of the locking driver delay time for both c-axis directed rod and plate samples, while such sharp steps are not found for cube shaped samples. These results demonstrate that the linear AFMR must be near the bottom of the spin wave manifold states before it is driven unstable so that ILMs appear in a true 3-D spin wave gap. The emission ladders observed for the c-axis directed rod and plate samples are interpreted as successive unlocking of the individual ILMs from the driver. In one kind of experiment where $\Delta f$ is varied as a function of the AFMR recovery, not only are steps observed but also an intriguing pattern of missing emission data appears in certain regions of parameter space. The traces have now coalesced into families. The measured values for the uniformly distributed case and the case with families are different. Similarly there are other regions where the emission signal is evident but steps are not found even though the locking frequency or the power has been varied. There is both simplicity and complexity to these observed time dependent emission spectra.

The locking time delay experiments provide a different way to examine these emission steps. Now at a long time (~ 6 million ILM periods) $\Delta f$ is essentially fixed and the number of locked ILMs is varied by interrupting the locking driver at a very early time. As a function of this delay the emission steps are well defined and the steps of individual scans appear at the same delay time with only a very small amount of time jitter. Since the step heights do not change significantly as the power of the locking driver is changed, a remaining puzzle is that the experimentally observed step height doesn't show the expected large frequency dependence obtained from numerical simulations. This robustness of the emission steps against external perturbations leads us to propose that these excitations are displaying a discrete well-defined character. Because the step heights for a particular experiment are well defined and uniform we can estimate both the size of an ILM and its energy. The estimated size is almost an order of magnitude larger than the size obtained from numerical simulations. These experiments identify a new direction in nonlinear nano-science with the next experimental goal to launch and receive these localized energy 'hot spots' across a measurable distance.

**Acknowledgments**

This work is supported by NSF-DMR Grant No. 0301035 and by the Cornell Center for Materials Research under DMR-0079992

# Appendix A: The 3rd Order Nonlinear Magnetization of an Easy Axis Antiferromagnet

The sample is a biaxial antiferromagnet. Here, we review the four wave mixing for the simpler case of an easy z-axis antiferromagnet.[29,30] Writing the torque equation for the uniform mode in circularly polarized modes in the usual way and then taking the next time derivation one obtains the nonlinear equation of motion

$$\frac{d^2\hat{s}^\pm_{A,B}(t)}{dt^2} + \omega_0^2 \hat{s}^\pm_{A,B}(t) = \omega_0^2 \hat{s}^\pm_{A,B}(t)\hat{s}^\mp_{A,B}(t)\hat{s}^\pm_{A,B}(t), \tag{A1}$$

where the (+) sign identifies one circularly polarized mode and the (-) sign the other for each of the sublattices A and B. Here

$$\hat{s}^\pm_{A,B}(t) = \frac{1}{2}\left[\hat{s}^\pm_{A,B} e^{i\omega t} + (\hat{s}^\pm_{A,B})^* e^{-i\omega t}\right] \tag{A2}$$

with $\hat{s}^\pm_{A,B}$ the time independent complex amplitudes for each circularly polarized modes and $\omega_0$ is the uniform mode frequency. Equation (A1) has the same form as the nonlinear anharmonic oscillator equation in Ref. [28].

Considering the right hand side of Eq. (A1) as the generator of the 3rd order signals, the nonlinear response of the left hand side can be calculated. For oscillating components $\hat{s}^\pm_{A,B}$ at frequencies $\omega_2$ and $\omega_3$ in the spin motion, the right hand side (RHS) of the nonlinear equation (A.1) will have the components

$$[RHS]^{\pm}_{A,B} = \omega_0^2 \frac{1}{8} \left[ \hat{s}^{\pm}_{A,B}(\omega_2)e^{i\omega t} + \hat{s}^{\pm *}_{A,B}(\omega_2)e^{-i\omega t} + \hat{s}^{\pm}_{A,B}(\omega_3)e^{i\omega t} + \hat{s}^{\pm *}_{A,B}(\omega_3)e^{-i\omega t} \right]$$

$$\times \left[ \hat{s}^{\mp}_{A,B}(\omega_2)e^{i\omega t} + \hat{s}^{\mp *}_{A,B}(\omega_2)e^{-i\omega t} + \hat{s}^{\mp}_{A,B}(\omega_3)e^{i\omega t} + \hat{s}^{\mp *}_{A,B}(\omega_3)e^{-i\omega t} \right] \quad (A3)$$

$$\times \left[ \hat{s}^{\pm}_{A,B}(\omega_2)e^{i\omega t} + \hat{s}^{\pm *}_{A,B}(\omega_2)e^{-i\omega t} + \hat{s}^{\pm}_{A,B}(\omega_3)e^{i\omega t} + \hat{s}^{\pm *}_{A,B}(\omega_3)e^{-i\omega t} \right]$$

Picking only terms proportional to $\exp[i(2\omega_2 - \omega_3)t]$ gives

$$[RHS]^{\pm}_{A,B}(t) = \omega_0^2 \frac{1}{8} \left[ 2\hat{s}^{\pm}_{A,B}(\omega_2)\hat{s}^{\mp}_{A,B}(\omega_2)\hat{s}^{\pm *}_{A,B}(\omega_3) + \hat{s}^{\pm}_{A,B}(\omega_2)\hat{s}^{\mp *}_{A,B}(\omega_3)\hat{s}^{\pm}_{A,B}(\omega_2) \right]$$
$$\times \exp[i(2\omega_2 - \omega_3)t] + c.c. \quad (A4)$$

The left hand side of Eq. A1 at the same frequency is

$$[LHS]^{(3)\pm}_{A,B}(t) = D(2\omega_2 - \omega_3)\frac{1}{2}\hat{s}^{(3)\pm}_{A,B}\exp[i(2\omega_2 - \omega_3)t] + c.c., \quad (A5)$$

where $D(\omega) = (-\omega^2 + \omega_0^2)$. From Eqs. (A4) and (A5), the nonlinearly generated transverse spin amplitude becomes

$$\hat{s}^{(3)\pm}_{A,B} = \frac{1}{D(2\omega_2 - \omega_3)}\omega_0^2\frac{1}{4}\left[ 2\hat{s}^{\pm}_{A,B}(\omega_2)\hat{s}^{\mp}_{A,B}(\omega_2)\hat{s}^{\pm *}_{A,B}(\omega_3) + \hat{s}^{\pm}_{A,B}(\omega_2)\hat{s}^{\mp *}_{A,B}(\omega_3)\hat{s}^{\pm}_{A,B}(\omega_2) \right]. \quad (A6)$$

Next we consider the nonlinear response function along the x direction for an ac magnetic fields applied along x direction. The net transverse magnetization along the x axis is

$$\hat{M}^x = \hat{M}^x_A + \hat{M}^x_B = \hat{\chi}^x \hat{H}^x \tag{A7}$$

where $\hat{\chi}^x$ is the susceptibility and $\hat{H}^x$ is the magnetic field along x direction. The transverse magnetization contribution from each sublattices is

$$\hat{M}^x_{A,B} = \hat{s}^x_{A,B} S g \mu_B N / 2 = \hat{s}^x_{A,B} M_0 \tag{A8}$$

where g is the g factor, $\mu_B$ is the Bohr magneton and N is the spin volume density. For the linear polarized field $\hat{H}^x$, the circularly polarized modes of equal amplitude are excited so

$$\hat{s}^+_A = \hat{s}^+_B = \hat{s}^-_A = \hat{s}^-_B = \frac{1}{4M_0} \hat{\chi}^x \hat{H}^x. \tag{A9}$$

Replacing spin amplitudes in Eq. (A6) with the fields in Eq. (A9) gives

$$\hat{s}^{(3)\pm}_{A,B} = \frac{1}{D(2\omega_2 - \omega_3)} \omega_0^2 \frac{3}{256 M_0^3} \left[\hat{\chi}^x(\omega_2)\hat{H}^x(\omega_2)\right]^2 \hat{\chi}^{x*}(\omega_3)\hat{H}^{x*}(\omega_3) . \tag{A10}$$

Converting this to magnetization, and summing over both polarizations and both sublattices, gives the third order nonlinear transverse magnetization

$$\hat{M}^{(3)}(2\omega_2 - \omega_3) = \frac{1}{D(2\omega_2 - \omega_3)} \omega_0^2 \frac{3}{64 M_0^2} \left[\hat{\chi}^x(\omega_2)\hat{H}^x(\omega_2)\right]^2 \hat{\chi}^{x*}(\omega_3)\hat{H}^{x*}(\omega_3). \tag{A11}$$

We now want to recast the linear response function in terms of the static and dynamic susceptibility of the system so

$$\frac{1}{D(\omega)} = \frac{\hat{\chi}^x(\omega)}{\chi^{x\prime\prime}(\omega_0)\gamma\omega_0} = \frac{\hat{\chi}^x(\omega)}{\chi_\perp(0)\omega_0^2}. \tag{A12}$$

Inserting Eq. (A12) into Eq. (A11) gives

$$\hat{M}^{(3)}(2\omega_2 - \omega_3) = \frac{3}{64 M_0^2} \frac{\hat{\chi}^x(2\omega_2 - \omega_3)}{\chi_\perp(0)} \left[\hat{\chi}^x(\omega_2)\hat{H}^x(\omega_2)\right]^2 \hat{\chi}^{x*}(\omega_3)\hat{H}^{x*}(\omega_3). \tag{A13}$$

The final expression in terms of the linear transverse magnetization at that frequency is

$$\hat{M}^{(3)}(2\omega_2 - \omega_3) = \frac{3}{64 M_0^2} \frac{\hat{\chi}^x(2\omega_2 - \omega_3)}{\chi_\perp(0)} \left[\hat{M}^x(\omega_2)\right]^2 \hat{M}^{x*}(\omega_3). \tag{A14}$$

## Appendix B: Stored energy density expression for a driven uniform mode

The eigenvector for the driven uniform mode is shown in the insert in Fig. 2(a). The numbers "1" or "3" identify the spin configuration at one instant of time when both spins are canted in the polarization direction giving the net transverse magnetization $M_c$. Starting with the biaxial Hamiltonian based on Ref. [24], the excitation state energy can be obtained. Here, the axes definition is the same as in the classical spin section of Ref.[24], that is, (a,b,c) = (x,y,z) = (easy, 2nd easy, hard) axes.

From Eqs. (2) and (3) in Ref. [24], the Hamiltonian density at zero external field is

$$H = -\frac{1}{2}\mu_0\lambda\left[\left(\vec{M}_A\right)^2 + \left(\vec{M}_B\right)^2\right] + \mu_0\lambda'\vec{M}_A \cdot \vec{M}_B - \frac{1}{2}\mu_0\vec{M}_A\left(\vec{A}+\vec{\vec{D}}\right)\vec{M}_A$$
$$-\frac{1}{2}\mu_0\vec{M}_B\left(\vec{A}+\vec{\vec{D}}\right)\vec{M}_B - \mu_0\vec{M}_A\vec{\vec{E}}\vec{M}_B + \frac{1}{2}\mu_0\left(\vec{M}_A+\vec{M}_B\right)\vec{\vec{N}}\left(\vec{M}_A+\vec{M}_B\right) \quad \text{(B1)}$$
$$-\frac{1}{2}\mu_0\frac{1}{3}\left(\vec{M}_A+\vec{M}_B\right)^2,$$

where $\vec{M}_A$ and $\vec{M}_B$ are the magnetization of each sublatttice, $\lambda$ and $\lambda'$ are ferromagnetic and antiferromagnetic molecular field parameters, $\vec{\vec{N}}$ simplifies for the rod-like sample to $N_x = N_y = 1/2$ and $N_z = 0$. The anisotropy field components of $\vec{\vec{A}}$ are $\left(A_x = 0, A_y = -H_{A1}/M_0, A_z = -H_{A2}/M_0\right)$, similarly both $\vec{\vec{D}}$, and $\vec{\vec{E}}$ produce dipole-dipole interaction fields on the diagonal from the lattice sums for one and the other sublattice, respectively. Setting the ground state spin configurations to

$$\vec{M}_A = (M_0, 0, 0), \vec{M}_B = (-M_0, 0, 0). \tag{B2}$$

in Eq. (B1) gives the ground state energy density

$$U_{ground} = -\mu_0 A_x M_0^2 - \mu_0 D_x M_0^2 + \mu_0 E_x M_0^2 - \mu_0 \lambda M_0^2 - \mu_0 \lambda' M_0^2. \tag{B3}$$

The lower branch AFMR excitation pattern shown in the inset of Fig. 2(a) generates a net linearly polarized magnetization along c axis at the moment indicated by letter "3". This state is expressed as

$$\vec{M}^A = \left(\sqrt{M_0^2 - |\hat{M}_c|^2/4}, 0, |\hat{M}_c|/2\right), \vec{M}^B = \left(-\sqrt{M_0^2 - |\hat{M}_c|^2/4}, 0, |\hat{M}_c|/2\right), \tag{B4}$$

where the ac magnetization amplitude along the hard axis is $|\hat{M}_c|$. Replacing $\vec{M}_A$ and $\vec{M}_B$ in Eq. (B1) by the large amplitude expressions in Eq. (B4), gives an energy density

$$\begin{aligned} U = &-\mu_0 \lambda M_0^2 - \mu_0 \lambda' M_0^2 + \mu_0 \lambda' \frac{1}{2}|\hat{M}_c|^2 - \mu_0 (A_x + D_x) M_0^2 \\ &+ \mu_0 (A_x - A_z + D_x - D_z) \frac{1}{4}|\hat{M}_c|^2 + \mu_0 E_x M_0^2 - \mu_0 (E_x + E_z) \frac{1}{4}|\hat{M}_c|^2. \\ &+ \frac{1}{2}\mu_0 \left(N_z - \frac{1}{3}\right)|\hat{M}_c|^2 \end{aligned} \tag{B5}$$

Subtracting Eq. (B3) from Eq. (B5) gives the driven energy density with respect to the ground state energy, namely,

$$\Delta U(M_c) = \mu_0 \frac{|\hat{M}_c|^2}{4}\left(A_x - A_z + D_x - D_z - E_x - E_z + 2(N_z - \tfrac{1}{3}) + 2\lambda'\right)$$
$$= \mu_0 \frac{|\hat{M}_c|^2}{4M_0}\left(H_{A2} + 2(N_z - \tfrac{1}{3})M_0 + 2H_E' + (D_x - D_z - E_x - E_z)M_0\right) \quad (B6)$$
$$= \mu_0 \frac{|\hat{M}_c|^2}{4M_0} H_{c-eff}.$$

Here the antiferromagnetic exchange field $H_E' = \lambda' M_0$ and the effective field along the c direction (z axis) for this set of precessing spins $H_{c-eff}$ is defined by Eq. (B6). The value of $H_{c-eff}$ evaluated from Ref. [24] is

$$H_{c-eff} = 3160 [Oe] = 4.0 \times 10^4 [A/m]. \quad (B7)$$

TABLE I. Analysis of the emission step height results.  The data in the AFMR column is used to determine $\alpha$ in Eq. (6). ILM, case 1 is for analyzing the data in Fig. 9(b) and ILM, case 2 is for analyzing the data in Fig. 9(d) using Eqs. (10), (11) and (13).

| | AFMR | ILM case 1[*a] | ILM case 2[*b] |
|---|---|---|---|
| $H(\omega_2)$ [$A/m$] | 35 | 19 | 20 |
| $H(\omega_3)$ [$A/m$] | 2.5 | 2.5 | 2.5 |
| $\omega_2/2\pi$ [$GHz$] | 1.32 | 1.335 | 1.330 |
| $\omega_3/2\pi$ [$GHz$] | 1.36 | 1.330 | 1.325 |
| $(2\omega_2-\omega_3)/2\pi$ [$GHz$] | 1.28 | 1.340 | 1.335 |
| $P^{(3)}$ [$nW$] | 0.29 | - | - |
| $\Delta\sqrt{P^{(3)}}$ [$(nW)^{1/2}$] | - | 0.30 | 0.26 |
| $v_\ell/v$ | - | $4.1\times 10^{-5}$ | $3.2\times 10^{-5}$ |
| $\Delta L$ [AFM cell] | - | 54 | 42 |
| $N_{spin}$ | - | $7.1\times 10^{14}$ | $5.5\times 10^{14}$ |
| $\delta(\Delta E_c)$ [$J$] | - | $1.5\times 10^{-12}$ | $1.3\times 10^{-12}$ |
| $\delta(\Delta E_c)/N_{spin}$ [$J$] | - | $2.1\times 10^{-27}$ | $2.4\times 10^{-27}$ |

[*a]From Fig. 9(b).

[*b]From Fig. 9(d).

**Figure captions**

**Figure 1.** Lattice and spin structure of $(C_2H_5NH_3)_2CuCl_4$. Circles denote $Cu_2^+$ ions and arrows indicate spin configuration in the antiferromagnetic state. Only $Cu_2^+$ ions are shown in this layered, face centered, orthorhombic compound. The easy, 2nd easy and hard spin axes are labeled the a, b, and c crystal directions, respectively.

**Figure 2.** Spin wave dispersion curve of the antiferromagnet $(C_2H_5NH_3)_2CuCl_4$. (a) Upper and lower branches along the c-axis. The inset shows the uniform mode spin motion for the lower AFMR mode, which has a net ac magnetization only along the c axis. The stored energy density for this excitation is described in Appendix B. The axes are identified in Fig. 1. (b) Expanded view of the lower branch, near the zone center with dispersion curves now along all crystal axes. The AFMR frequencies for different sample shapes are indicated by the solid dots.

**Figure 3.** Frequency diagram for the four wave mixing experiment. Vertical arrows represent transverse ac magnetizations. The $f_2$ driver and $f_3$ probe produce transverse magnetizations $M_2$ and $M_3$ inside the sample. The sample then generates the new magnetization at frequency $(2f_2 - f_3)$ by the 3rd order nonlinearity, as indicated by the downward arrow. The microwave signal from this magnetization is observed in emission.

**Figure 4.** Experimental setup for the ILM locked state measurement. Three microwave sources are used: a high power pulse pump source for the initial excitation ($f_1$), a source for locking ($f_2$), which is followed by a switch (SW) and a middle power amplifier (AMP), and a low power probe source ($f_3$). The high power pulse microwave driver $f_1$ excites the sample, which is immersed in 1.2 K liquid helium. The middle power $f_2$ cw driver is employed to lock ILMs. Here one branch of the microwave signal goes to the sample via the directional coupler. Since the reflected $f_2$ signal is often larger than the maximum linear input of the spectrum analyzer, the other branch of the $f_2$ driver is fed into the spectrum analyzer to cancel it, thus avoiding a spurious mixing signal inside the spectrum analyzer. The $f_3$ probe is used both in absorption and in emission measurements.

**Figure 5.** Time dependent absorption spectrum showing the break up of the uniform mode induced by a strong microwave pulse at frequency $f_1$. Here $f_1 = 1.29$ GHz, the input power at the cryostat is 52 W and the pulse length is 3 $\mu s$. The power of the cw driver $f_2$ (dotted line) is typically 1000 times smaller that $f_1$.

**Figure 6.** Time dependent emission signal showing both locking and releasing of ILMs. (a) Single on-off-on driver sequence. (b) Double on-off-on sequence. The square waves on the lower part of panels show the on-off pattern of the $f_2$ driver. The horizontal dotted line indicates the frequency position of the $f_2$ driver. Shortly after the $f_1$ driver is shortly on at $t = 20 \mu s$, the emission signal appears. The locked ILMs can be maintained as long as the driver is on. Once turned off, most of

**ILMs are unlocked and no longer emit coherently. When the driver is again turned on, only some ILMs are re-locked.**

**Figure 7. At 2 ms after the $f_1$ pulse a snapshot of the simultaneous emission (solid curve) and absorption (dotted curve) spectra. For this time scale, the frequency resolution is ~ 100 kHz, and the raw emission spectra are shown. (a) $f_2$ = 1.34 GHz at a power of 51 mW. (b) $f_2$ = 1.32 GHz at a power of 240 mW. Note the AFMR frequency is pulled slightly to lower frequencies by $f_2$ at this power level. In case (b) the small number of locked ILMs is not apparent in absorption. A number of features are seen in emission. The strongest and the 2nd strongest peaks on either side of the driver are associated with the locked ILMs, resulting in a sideband pair in both figures. The 3rd strongest peak at the AFMR frequency is emission from the this uniform mode. The other half of the sideband pair appears as a small shoulder in (a) and as a small peak in (b).**

**Figure 8. Complex time dependent emission output for a near cubic shaped sample. Square root of the time dependent emission output as a function of time for different values of the cw $f_2$ power level. The 2.3% increments between curves vary the $f_2$ power from 32.4 to 81.3 mW. The driver frequency $f_2$ is fixed at 1.35 GHz. AFMR frequency of this sample is 1.395 GHz somewhat higher than that of rod shaped samples (1.375 GHz). Although some emission plateaus can be identified, the overall structures are very complex.**

Figure 9. Square root of the time dependent emission output versus time for a c-axis directed thin plate sample. (a) For fixed driving frequency $f_2 = 1.33$ GHz as a function of the cw $f_2$ power level from 52.5 to 105 mW, (b) For fixed $f_2$ power level 77.6 mW, as a function of its frequency from 1.325 to 1.335 GHz, (c) For fixed frequency $f_2 = 1.335$ GHz as a function of the cw power level from 34.7 to 87.1 mW, (d) For fixed $f_2$ power level of 55 mW as a function of its frequency from 1.33 to 1.34 GHz. The power increment between curves is 2.3%. The frequency interval between curves is 250 KHz.

Figure 10. Characterizing the time dependence of the four wave emission signals. The dotted traces in Fig. 9(a,b) are singled out for analysis. (a) Dotted curve in 9(a) for conditions $f_2$=1.33 GHz at 83.2 mW; time constant of the dotted curve =1.18 ms, step height = 0.26 (nW)$^{1/2}$. (b) Dotted curve in 9(b) for conditions $f_2$=1.33123 GHz at 77.6mW, time constant of the dotted curve =1.18ms, step height = 0.26 (nW)$^{1/2}$. In both frames the steps are superimposed on an exponential time dependent background.

Figure 11. Emission decay curve dependence on the $f_2$ driver delay time. After ~ $3 \times 10^6$ $f_2$ periods the driver is turned off and after a brief delay again turned on. Solid curve, driver unchanged. The dotted, and dot-dashed curves are for 10, and 50 μs delay times. Unlocked ILMs lose phase coherence, shift in frequency so their emission signal quickly disappears. When the driver is turned on again, a subset of the ILMs oscillating around the driver frequency are re-locked. Different ILM states can be produced without changing the $f_2$ driver frequency or its power.

Figure 12. Frames (a, b). Square root of the emission power observed at t = 4ms as a function of the driver delay time at 2 ms for the c-axis directed plate sample. The $f_2$ driver frequency and power for the solid curves are: (a) 1.335 GHz, 93.3 mW; (b) 1.32GHz, 191 mW. The distance between the horizontal dotted lines is 0.26 $(nW)^{1/2}$. The dot-dash curves in (a) and (b) are for slightly higher power: (a) 102 mW, and (b) 200 mW. Frame (c). Results for a c-axis directed rod sample measured at 1.31 GHz, 97.7 mW. The distance between these horizontal dotted lines is 0.18 $(nW)^{1/2}$. The dot-dash curve is for the higher power, 110 mW.

Figure 13. (a) Illustration of the classical anharmonic oscillator resonance response curve for soft anharmonicity. The frequency axis is measured with respect to the higher AFMR frequency so that as $\Delta f = f_{AFMR} - f_\ell$ increases a single step can occur. (b) The proposed step emission pattern can occur when the locked ILM frequency depends weakly on the number of ILMs in the 1-D lattice.

Figure 14. Step height as a function of the normalized frequency difference between the AFMR and the $f_2$ driver as determined from numerical simulations. The Sum of the Square of the Transverse Component (SSTC) for an ILM eigenvector (proportional to the square root of the emission) versus the normalized frequency gap is shown, The region between the two vertical arrows corresponds to the range

of the $f_2$ frequency scanning experiment presented in Fig. 9(b). These simulations predict a 12% increase in step height for this fractional increase while the experiment shows less than a 4% change.

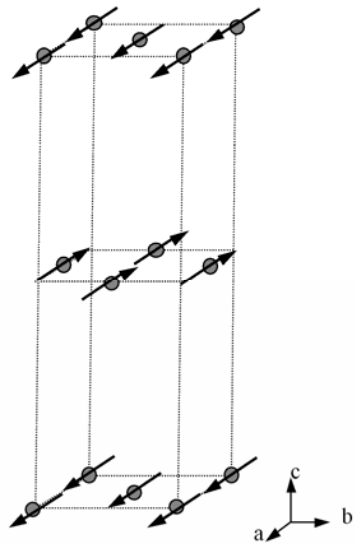

Figure 1.

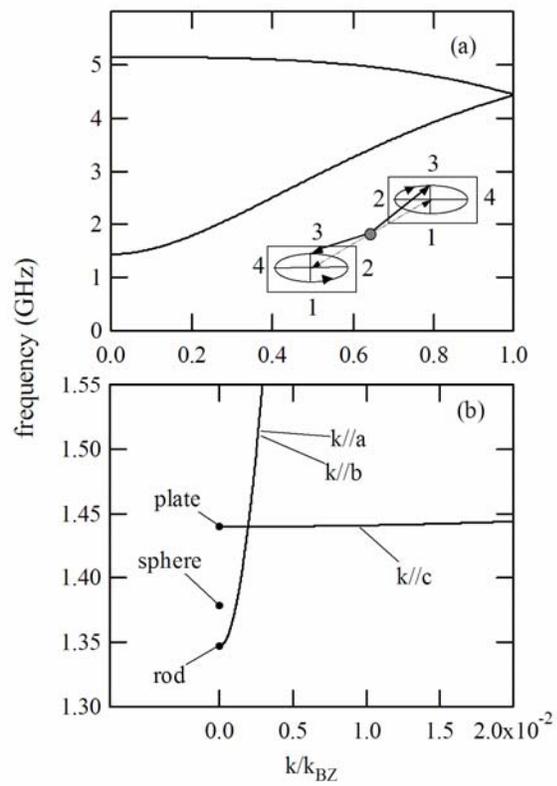

Figure 2.

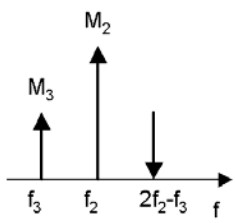

Figure 3.

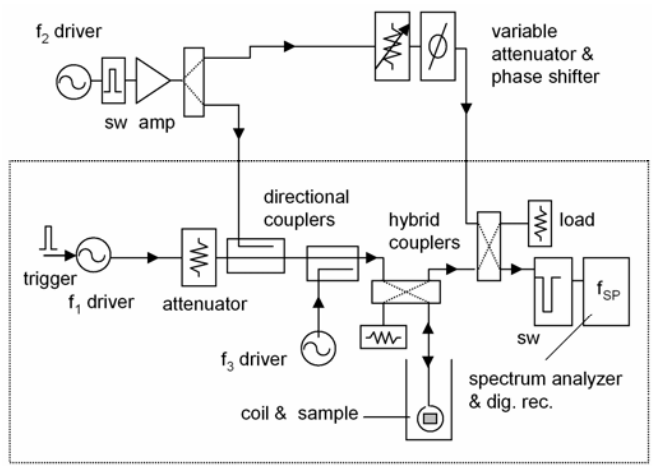

Figure 4.

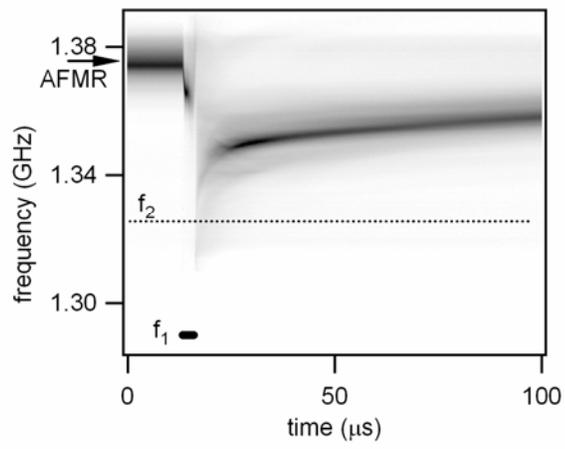

Figure 5.

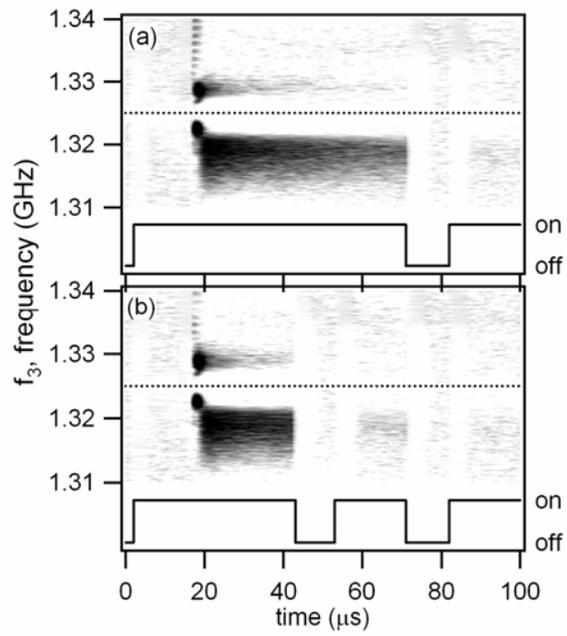

Figure 6.

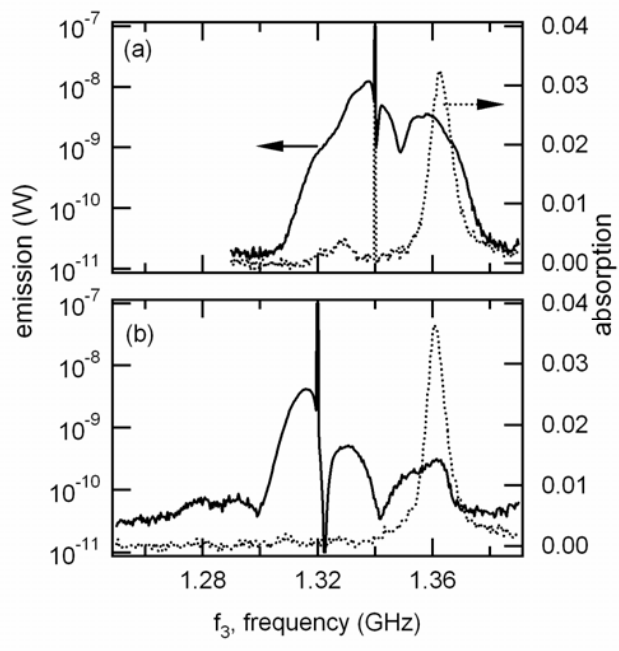

Figure 7.

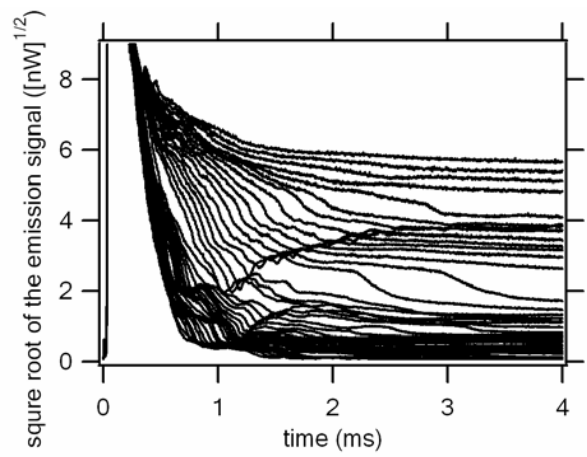

Figure 8.

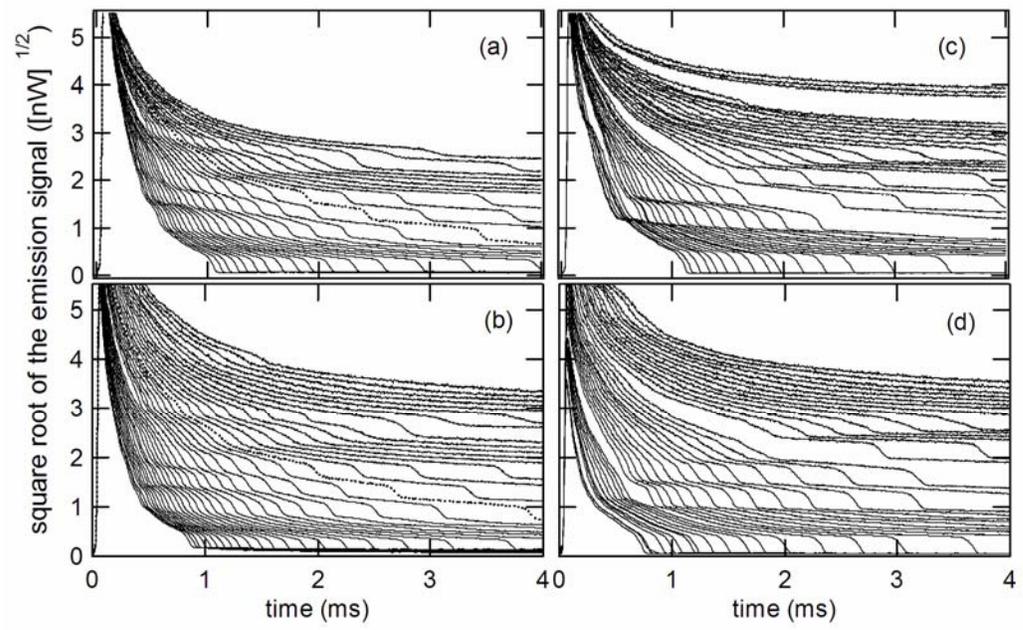

Figure 9.

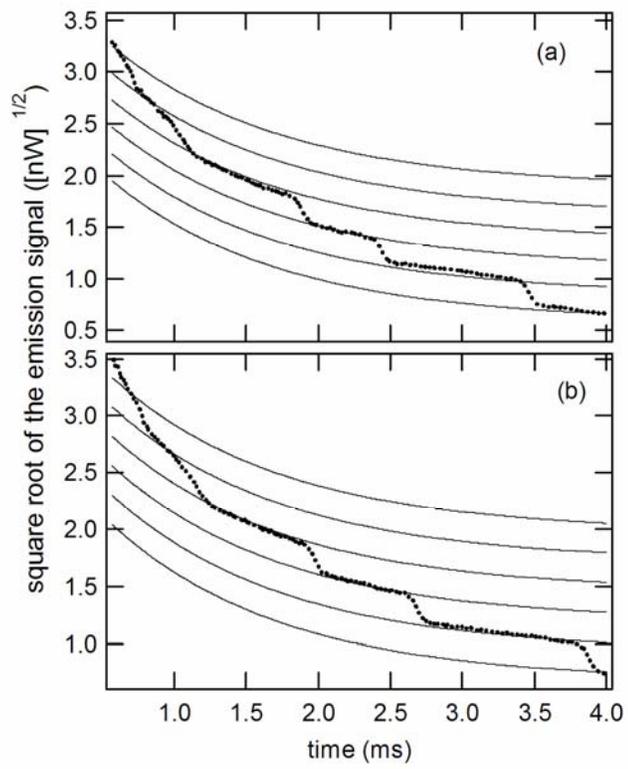

Figure 10.

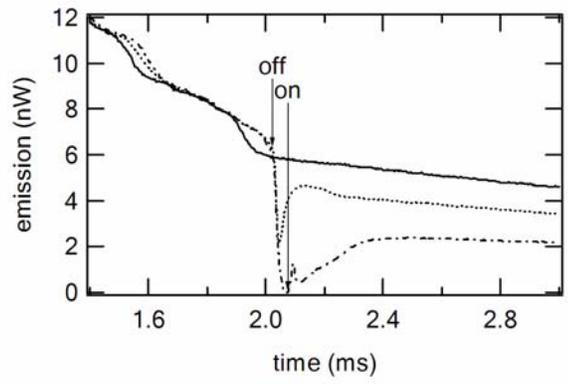

Figure 11.

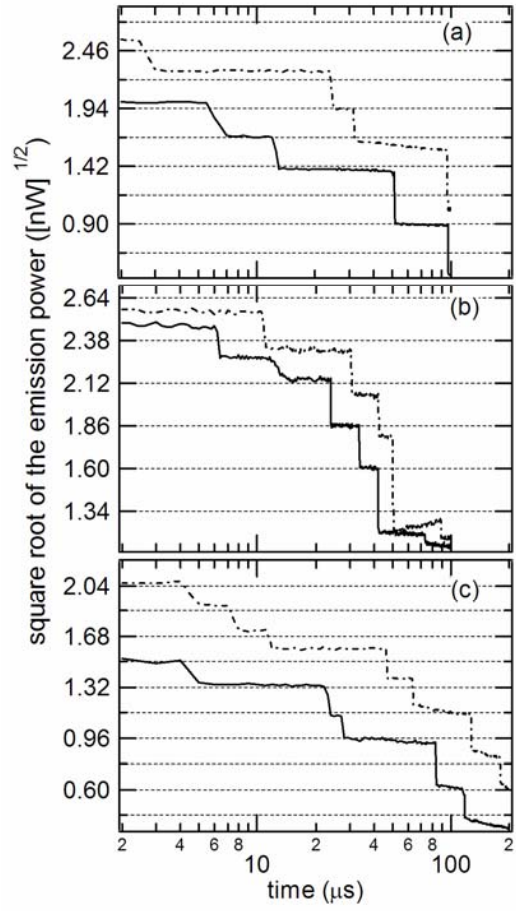

Figure 12.

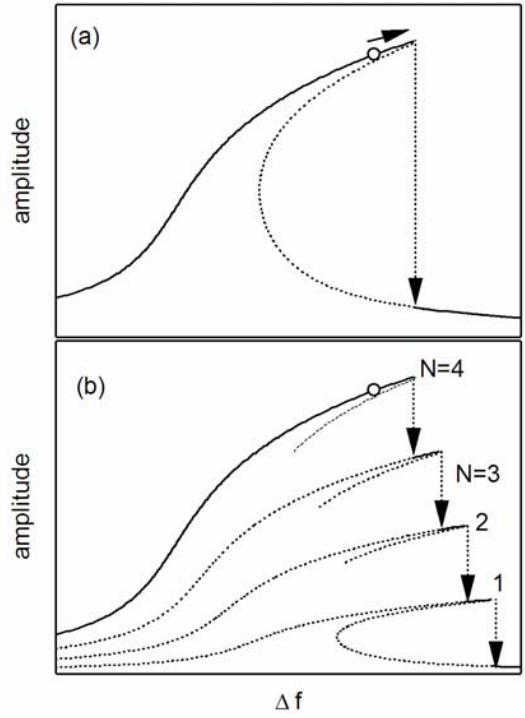

Figure 13.

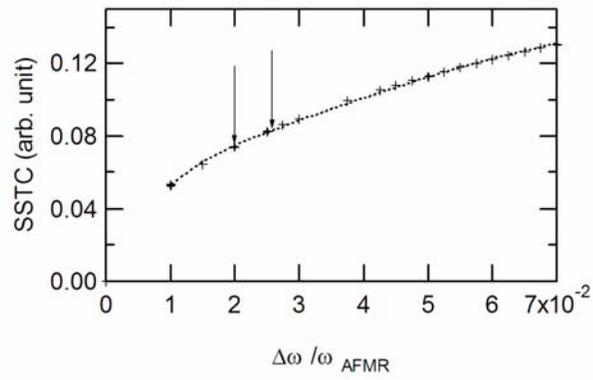

Figure 14.